# High Extinction Ratio Widely Tunable Low-Loss Integrated Si$_3$N$_4$ Third-Order Filter

**TARAN HUFFMAN[1], DOUG BANEY[2], DANIEL J. BLUMENTHAL*[1]**

[1]*Department of Electrical and Computer Engineering, University of California, Santa Barbara, CA 93106, USA*
[2]*Keysight Laboratories, Keysight Technologies, 5301 Stevens Creek Blvd., Santa Clara, CA 95051, USA*
[*danb@ece.ucsb.edu](danb@ece.ucsb.edu)

**Abstract:** We demonstrate an integrated continuously tunable third-order ring filter with a measured extinction ratio of 80dB, a 100x improvement over previously demonstrated integrated 3$^{rd}$ or 6$^{th}$ order filters.  Using integrated thermal tuning elements, the filter can be tuned over 100% of its 48 GHz free spectral range with 1.3 dB insertion loss. The filters are fabricated on a wafer-scale foundry compatible Si$_3$N$_4$ low-loss platform enabling integration with a wide variety of previously demonstrated passive and active elements.  The high extinction ratio, low loss, flat passband and steep roll-off is desirable for a broad range of applications including pump-stokes separation for Brillouin scattering, idler separation in four-wave mixing and separation of entangled states for quantum communications that utilize nonlinear optics.

## References

1. Kittlaus, Eric A., Heedeuk Shin, and Peter T. Rakich. "Large Brillouin amplification in silicon." *Nature Photonics* (2016).
2. Hua, Shiyue, Jianming Wen, Xiaoshun Jiang, Qian Hua, Liang Jiang, and Min Xiao, "Demonstration of a chip-based optical isolator with parametric amplification," *Nature Communications* | 7:13657 | DOI: 10.1038/Ncomms13657
3. Silverstone, Joshua W., Bonneau D, Ohira K, Suzuki N, Yoshida H, Iizuka N, Ezaki M, Natarajan CM, Tanner MG, Hadfield RH, Zwiller V. "On-chip quantum interference between silicon photon-pair sources." *Nature Photonics* 8.2 (2014): 104-108.
4. Popovíc, Miloš A., Tymon Barwicz, Michael R. Watts, Peter T. Rakich, Luciano Socci, Erich P. Ippen, Franz X. Kärtner, and Henry I. Smith.  "Multistage high-order microring-resonator add-drop filters." *Optics Letters* 31.17 (2006): 2571-2573.
5. Smith, Henry, Tymon Barwicz, Charles W. Holzwarth, Milos A. Popovic, Michael R. Watts, Peter T. Rakich, Minghao Qi, Raul Barreto, Franz X. Kärtner, and Erich P. Ippen. "Strategies for fabricating strong-confinement microring filters and circuits." *Optical Fiber Communication Conference*. Optical Society of America, 2007.
6. Little, Brent. "Advances in microring resonators." *Integrated Photonics Research*. Optical Society of America, 2003.
7. Little, B. E., Chu, S.T., Absil, P.P., Hryniewicz, J.V., Johnson, F.G., Seiferth, F., Gill, D., Van, V., King, O. and Trakalo, M. "Very high-order microring resonator filters for WDM applications." *IEEE Photonics Technology Letters* 16.10 (2004): 2263-2265.
8. Little, Brent, Chu, S., Chen, W., Hryniewicz, J., Gill, D., King, O., Johnson, F., Davidson, R., Donovan, K., Chen, W. and Grubb, S. "Tunable bandwidth microring resonator filters." *Optical Communication, 2008. ECOC 2008. 34th European Conference on*. IEEE, 2008.
9. Dong, Po, Feng, N.N., Feng, D., Qian, W., Liang, H., Lee, D.C., Luff, B.J., Banwell, T., Agarwal, A., Toliver, P. and Menendez, R. "GHz-bandwidth optical filters based on high-order silicon ring resonators." *Optics Express* 18.23 (2010): 23784-23789.
10. Orlandi, P., Morichetti, F., Strain, M. J., Sorel, M., Bassi, P., & Melloni, A.  "Photonic integrated filter with widely tunable bandwidth." *Journal of Lightwave Technology* 32, no. 5 (2014): 897-907.
11. Ong, Jun Rong, Ranjeet Kumar, and Shayan Mookherjea. "Ultra-high-contrast and tunable-bandwidth filter using cascaded high-order silicon microring filters." *IEEE Photonics Technology Letters* 25.16 (2013): 1543-1546.
12. Jared F Bauters, Martijn JR Heck, Demis John, Daoxin Dai, Ming-Chun Tien, Jonathon S. Barton, Arne Leinse, René G. Heideman, Daniel J. Blumenthal, and John E. Bowers. "Ultra-low-loss high-aspect-ratio Si 3 N 4 waveguides." *Optics express* 19, no. 4 (2011): 3163-3174.
13. A. Leinse, Shaoxian Zhang and R. Heideman, "TriPleX: The versatile silicon nitride waveguide platform," *2016 Progress in Electromagnetic Research Symposium (PIERS)*, Shanghai, 2016, pp. 67-67.

14. Chaichuay, Chinda, Preecha P. Yupapin, and Prajak Saeung. "The serially coupled multiple ring resonator filters and Vernier effect." *Opt. Appl* 39.1 (2009): 175-194.
15. Belt, Michael, and Daniel J. Blumenthal. "Erbium-doped waveguide DBR and DFB laser arrays integrated within an ultra-low-loss Si 3 N 4 platform." *Optics Express* 22.9 (2014): 10655-10660.
16. Piels, M., Bauters, J. F., Davenport, M. L., Heck, M. J., & Bowers, J. E. (2014). Low-loss silicon nitride AWG demultiplexer heterogeneously integrated with hybrid III–V/silicon photodetectors. *Journal of Lightwave Technology*, *32*(4), 817-823.
17. Moreira, Renan L., John Garcia, Wenzao Li, Jared Bauters, Jonathon S. Barton, Martijn JR Heck, John E. Bowers, and Daniel J. Blumenthal. "Integrated ultra-low-loss 4-bit tunable delay for broadband phased array antenna applications." *IEEE Photon. Technol. Lett* 25.12 (2013): 1165-1168.
18. Rabus, Dominik G. *Integrated ring resonators*. Springer-Verlag Berlin Heidelberg, 2007.
19. Huffman, T., Davenport, M., Belt, M., Bowers, J. E., & Blumenthal, D. J. (2016). Ultra-low loss large area waveguide coils for integrated optical gyroscopes. *IEEE Photonics Technology Letters*, *29*(2), 185-188.
20. Sudbo, A. Sv. "Film mode matching: a versatile numerical method for vector mode field calculations in dielectric waveguides." *Pure and Applied Optics: Journal of the European Optical Society Part A* 2.3 (1993): 211.
21. Amatya, Reja, Charles W. Holzwarth, M. A. Popovic, F. Gan, H. I. Smith, F. Kartner, and R. J. Ram. "Low power thermal tuning of second-order microring resonators." *Conference on Lasers and Electro-Optics*. Optical Society of America, 2007.
22. Mak, J. C., Sacher, W. D., Xue, T., Mikkelsen, J. C., Yong, Z., & Poon, J. K. (2015). Automatic resonance alignment of high-order microring filters. *IEEE Journal of Quantum Electronics*, *51*(11), 1-11.

## 1 Introduction

Widely tunable filters with very high extinction ratios, low loss, and flat passbands are important for many applications including communications, lasers, spectroscopy and nonlinear optics. Examples include separating pump and Stokes signals for Brillouin scattering [1], filtering of idler signals in FWM processes for nonlinear micro-resonators and non-magnetic optical isolation [2], and quantum communications and computing that employ frequency conversion [3].

Integrated planar waveguide coupled-ring structures have been shown to realize filters with reduced cost and fabrication complexity and scaling to larger more complex PIC circuits. To date multiple-order non-tunable ring filters have been demonstrated with extinction ratios of up to 70dB. Popovíc, Miloš A., et al. [4] and Smith, Henry, et al. [5] demonstrated high confinement silicon rich $Si_xN_4$ third-order ring filters with 50dB and 60dB extinction ratios, respectively, and FSRs up to 2500GHz. Little, et al. demonstrated 3rd order filters with 50dB extinction ratio with 3dB bandwidth between 10-15GHz and 6th order filters with extinction ratio on the order of 60dB [6,7], with a discretely tunable passband shape [8]. In addition, Little presented results on an 11th order filter with asymmetric shape that experimentally demonstrated 50 dB symmetric extinction ratio and approximately 70dB asymmetric extinction ratio without clear measurement of the filter stopband [7] and discrete tuning over an unspecified percentage of the FSR. Dong, Po, et al. demonstrated silicon 5th order tunable ring filters with up to 50dB extinction and an FSR of 50GHz [9]. P. Orlandi et al. utilized a ring-loaded MZI geometry to demonstrate 10dB − 45dB ER with tuning ranges up to 90% of 200GHz FSR with a 20dB extinction ratio [10]. Jung Rong Ong reported a 60dB ER 5th order filter using silicon microrings, that demonstrated a shapeable passband and was tunable over a small portion of its FSR [11].

In this work, we report the demonstration of a widely tunable third-order ring filter with 80dB extinction ratio, a 100x improvement in extinction ratio over prior art 3rd and 6th order filters [6] and 3000x larger tuning range than reported in [10]. The filter is continuously tunable over 100% of the free spectral range using the thermo-optic effect with a measured insertion loss of 1.3dB. The filters are fabricated using the $Si_3N_4$ ultra-low loss waveguide platform (ULLW) reported in [12] and are compatible with the wafer-scale foundry processes described in [13]. This advance in state of the art is due to the low waveguide propagation losses and reproducible and accurate waveguide coupling regions as well as the ability to design and implements large FSR tunable structures. Tuning is achieved with monolithically integrated waveguide heaters for each ring. We use a unit-less quantity called the filter shape factor (SF) defined as the ratio of the -1dB and -10dB bandwidths [14] to evaluate the filter passband roll off. We report a measured shape factor better than 0.4 that is maintained while tuning over the full 48.2GHz FSR. Implementation of this filter in the $Si_3N_4$ platform lends to monolithic integration with a wide range of both active and passive components that have been previously demonstrated, including $Si_3N_4$-core waveguides with co-doped $AL_2O_3$:$Er^{3+}$ doped narrow linewidth WDM sources [15], AWGRs [16] and thermal switches and delay lines [17].

## 2   Integration Platform

The filter consists of three coupled-ring waveguide resonators interfaced via directional couplers to input and output bus waveguides as illustrated in Fig. 1 (a).  S-bends to the directional couplers are used to isolate the fiber-coupled input and output buses from the ring resonators.  The low loss waveguides, couplers, and rings are fabricated in a common planar wafer-scale compatible structure illustrated in Fig. 1 (b) that incorporates a thin $Si_3N_4$ core, a thermally grown $SiO_2$ on silicon lower cladding and a PECVD deposited upper cladding layer with a metal heater layer deposited on top to thermally tune the rings.

The waveguide propagation loss components are defined by sidewall scattering at the core-cladding interface, the waveguide bend radius, and the material losses at the operational wavelengths.  The sidewall scattering loss scales quadratically with the interface roughness and dominates the waveguide losses for bends larger than the critical loss radius.  We employ high-aspect ratio waveguides (t1<<w) to reduce the scattering loss of the etched waveguide sidewalls [12].  This large aspect ratio causes the TE mode to be more highly confined than the TM mode allowing the TE mode to propagate in smaller radius bends than the TM mode before experiencing significant bend loss.

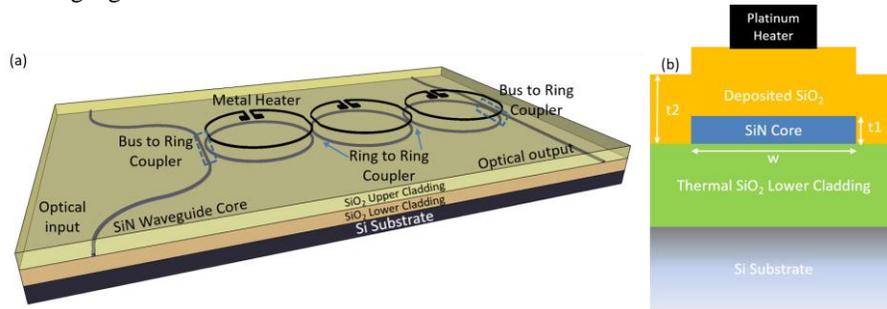

Fig. 1.    (a) Third-order ring filter design; (b) Schematic cross section of $Si_3N_4$ low loss waveguide.  In this device, we use nitride core thickness t1 = 175nm, core width w = 2.2µm and upper cladding thickness t2 =6.8µm. Thermal oxide lower cladding thickness is 15 microns.

## 3   Third-Order Filter Design

The basic filter design and parameters are shown in Fig. 2. The coupling between rings and the bus waveguides, the ring radius, and the waveguide loss determine the performance and shape of the filter including extinction ratio, shape factor, and ripple.

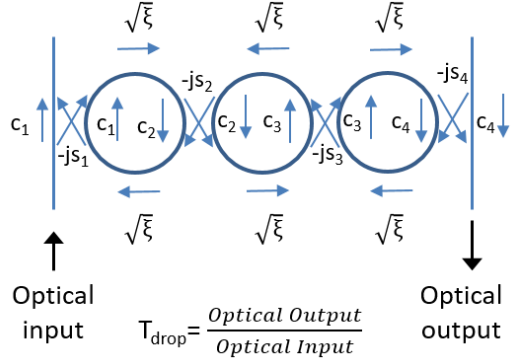

Fig. 2. Bus coupled ring-resonator third-order filter design and parameters.

The transfer function of the drop port $T_{drop}$, designated as the ratio of the output optical power to the input optical power, is defined using Mason's rule [14] and given in Eq. (1). Light passing through couplers without coupling is represented by the parameter $C_i$, while light coupling between waveguides is represented by the parameter $S_i$, and the attenuation and phase of light traveling around a ring is represented by the parameter $\xi$. These parameters are defined in Eqns. (2-4) and are illustrated in Fig. 2. Here L is the round-trip length of the ring, $\alpha$ is the propagation loss of the waveguide, and $\beta$ the waveguide propagation constant.

$$T_{drop} = \frac{S_1 S_2 S_3 S_4 \xi^{3/2}}{1 - C_1 C_2 \xi + C_1 C_3 \xi^2 + C_2 C_4 \xi^2 - C_1 C_4 \xi^3 + C_1 C_2 C_3 C_4 \xi^2} \quad (1)$$

$$C_i = \left((1-\kappa_i)(1-\gamma)\right)^{1/2} \quad (2)$$

$$-jS_i = -\left((1-j\gamma)\kappa_i\right)^{1/2} \quad (3)$$

$$\xi = e^{\alpha L/2} e^{-j\beta L} \quad (4)$$

The shape factor is defined as the ratio of bandwidth at the -1dB and -10dB points where a shape factor of 1 is a perfect box-like filter function. For a given radius and propagation loss, the ratio of $\kappa_1$ and $\kappa_2$ determine the ripple, insertion loss (IL) and extinction ratio (ER). To minimize the filter IL, the bus to ring coupling constants are set to $\kappa_1 = \kappa_4$ and ring to ring coupling constants $\kappa_2 = \kappa_3$ as described in [18]. High ratio values of $\kappa_2$ to $\kappa_1$ will produce low ER and high ripple whereas low values will decrease the shape factor and increase insertion loss. A maximally flat filter shape is derived for the lossless case given in [18] as $\kappa_1^2 = .125 * \kappa_2^4$. Changing $\kappa_1$ and $\kappa_2$ from this ideal case allows the ER and IL of the filter to be varied.

Fig. 3 (a) shows the calculated drop port transfer function with an ideal coupling ratio, Fig. 3 (b) shows the drop port transfer function with a low coupling ratio, and Fig. 3 (c) shows the drop port transfer function with a high coupling ratio. In these plots and all plots throughout this paper, the frequency dependent filter transmission is relative to a 1550nm center wavelength.

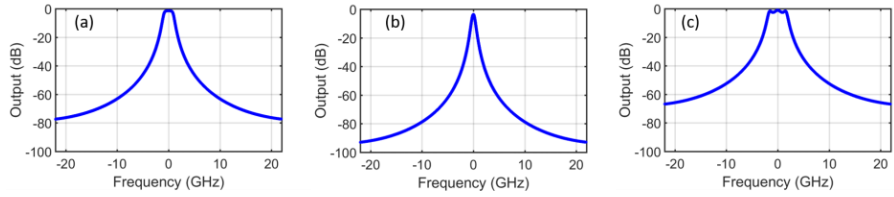

Fig. 3. Calculated drop port transfer function, (a) Shows an ideal coupling ratio filter with $\kappa_2$=0.006, yielding SF=0.6, ER=77dB, and IL=1.4dB, (b) Shows an under coupled filter with $\kappa_2$=0.001, yielding SF=.26, ER=90, and IL=3.6, (c) Shows an over coupled filter with $\kappa_2$=0.01, yielding SF=0.82, ER=67dB, IL=1dB, and ripple =2dB. All filters have a radius of 580μm, $\kappa_1$ of 0.15, and loss of 20dB/m.

## 4    Filter Design

### 4.1   Waveguide Geometry

An important design space is the tradeoff between filter performance and filter area. These factors are set by the waveguide propagation loss and minimum bend radius as determined by the $Si_3N_4$ core thickness and waveguide width. Thicker cores have higher sidewall scattering loss but a lower critical bend radius while thinner cores can greatly reduce the scattering propagation loss but result in a larger critical bend radius. The critical bend radius is defined as where the bend loss contribution is equal to 0.1dB/m. Table 1 summarizes experimental and simulated propagation loss values and bend limits for different core thicknesses as reported in [19,12]. In this work, the filter was designed for a 50GHz FSR and based on the parameters in Table 1, a core thickness of 175nm was selected. Waveguide mode intensity profiles for both the TE and TM modes in a 175nm thick core design were simulated using FIMMWAVE and the film mode matching technique [20] and are shown in Fig. 4. For this design, we chose TM operation due to initial fabrication runs that resulted in high interface roughness between the core and upper cladding from our sputtered upper cladding deposition process. The higher than expected interfaced roughness resulted in unacceptable TE losses. Using a PECVD deposition process for the upper cladding deposition reduced the TE mode loss, and this is reported for the results in section 6. Designing the filter for the TE mode using low loss deposition would allow reduction of the bend radius and therefore an increase in the FSR and decrease the round-trip loss.

**Table 1. Filter FSR and waveguide loss and critical bend radius dependence on core thickness**

| Core Thickness (nm) | Scattering Loss (dB/m) | Bend Limit (mm) | FSR Limit (GHz) |
|---|---|---|---|
| 40 | 0.2 — 0.5 (TE) | 11 | 2.93 |
| 60 | 0.8 — 3 (TE) | 3 | 10.7 |
| 90 | 1 — 6 (TE) | 1 | 32.3 |
| 175 (this work) | 10 — 20 (TE & TM) | .5 (TM) | 63.6 |

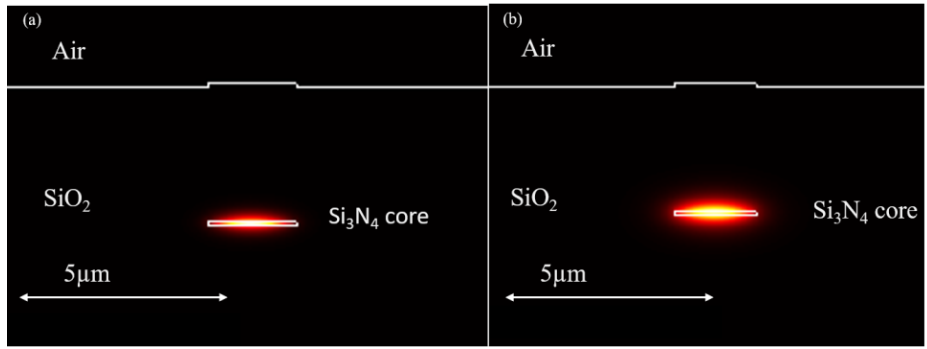

Fig. 4. Simulated mode shapes for both the (a) TE and (b) TM mode for a 175nm thick and 2.2um wide $Si_3N_4$ core.

### 4.2 Coupler Design

The coupling coefficients for the bus to ring and ring to ring couplers, $\kappa_1$ and $\kappa_2$, have a significant impact on filter performance. Simulating directional couplers with the precision needed to engineer a high extinction ratio third-order filter design requires calibration with the fabrication process. To determine the precise relationship between the waveguide coupling gaps and coupling coefficients, a coupling parameter split test run was fabricated and measured. We then used measured results of the fabricated first-order filter to design the third-order filter.

A first-order ring filter test structure, shown in Fig. 5, was used to calibrate the coupling coefficients to the coupling waveguide gap. The drop-port characteristics were measured with a laser wavelength sweep into a photodetector. The resulting filter shape was then fit to the equation for the drop port characteristic of a first-order ring in [14] using a least-squares fit, as shown in Fig. 6 (a). This yields the relationship between coupling gap and coupling coefficient, shown in Fig. 6 (b) for the TM mode.

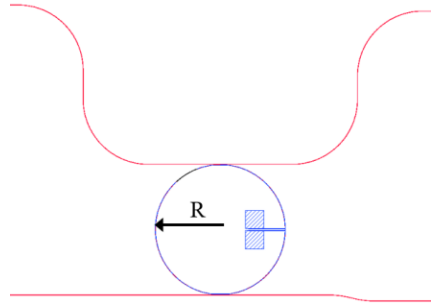

Fig. 5. Thermally tuned single order ring filter test structure with R=625μm.

The ratio between κ1 and κ2 was chosen for a target ER of 80 dB and a flat passband. The physical gap values used on the mask, and the corresponding intended coupling coefficients from Fig. 6 (b), for the third-order rings are summarized in Table 2.

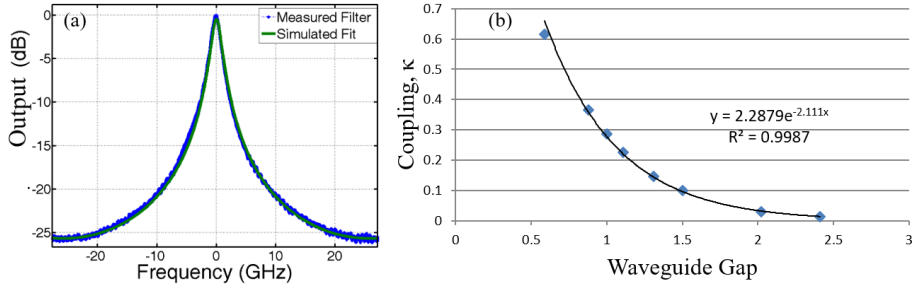

Fig. 6.  (a) The analytic fit to a first order ring filter and (b) the measured trend across multiple radii and gaps for the TM mode.

**Table 2. Gap values with corresponding Predicted Kappa Values**

|  | Gap (µm) | Kappa |
| --- | --- | --- |
| **Bus Coupler** | 1.15 | .13 |
| **Ring to Ring Coupler** | 2.4 | .06 |

### 4.3  Heater Layer and Upper Cladding Thickness

$Si_3N_4$ waveguides can be thermally tuned [21] using a resistive metal layer on the upper cladding over the core. The heaters are an absorptive metal layer deposited directly over the waveguide requiring careful selection of the upper cladding thickness, balancing excess loss from optical mode overlap with the heater and heater power tuning efficiency. Fig. 7 shows FIMMWAVE simulations of the estimated optical loss as a function of the upper cladding thickness. Modes with upper claddings thicker than 6.5µm will experience negligible loss from the metal layer.

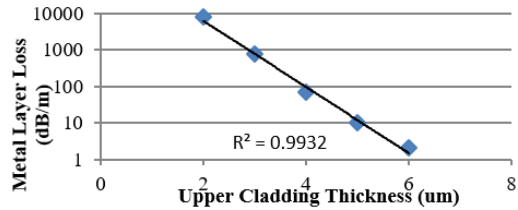

Fig. 7.  Metal absorption loss as a function of upper cladding thickness for the TM mode.

### 5  Fabrication Process

Fabrication begins with a 1mm thick Si substrate wafer with 15µm of thermally grown $SiO_2$. The 175nm core is deposited using low-pressure chemical vapor deposition (LPCVD) performed by Rogue Valley Microdevices. The core is defined using DUV lithography and a dry etch consisting of $CHF_3$, $CF_4$, and $O_2$. The upper cladding is deposited using plasma-enhanced chemical vapor deposition (PECVD). The deposition is performed in 3.4µm steps and annealed for 7 hours at 1050º C after each deposition. The process flow is shown in Fig. 8 processing quantities summarized in Table 2   Further details on the fabrication process can be found in [12].

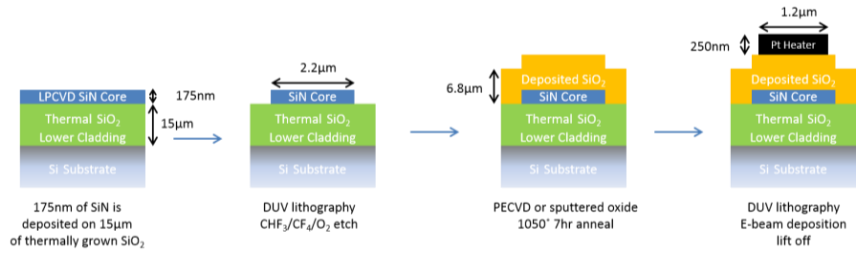

Fig. 8.    Cross-section schematic of process flow.

**Table 3. Process parameters**

| Parameter | Value |
| --- | --- |
| Lower Cladding Thermal $SiO_2$ Thickness | 15µm |
| Lower Cladding Thermal $SiO_2$ Index | 1.445 |
| $Si_3N_4$ Core Thickness | 173.9nm |
| $Si_3N_4$ Core Index | 1.983 |
| Core Etch Depth | 206nm |
| Upper Cladding PECVD $SiO_2$ Thickness | 6.8µm |
| Upper Cladding PECVD $SiO_2$ Index | 1.456 |
| Ti Thickness | 10nm |
| Pt Thickness | 250nm |

The metal heater layer is added using a lift-off technique and the same DUV stepper. It is relevant that a 1mm thick Si wafer is used, as a 0.5mm thick wafer will be rejected from the stepper due to bowing from the thick $SiO_2$ layer. The metal is deposited using e-beam evaporation. A small 10nm Ti layer is added first for adhesion, then the 250nm Pt heater itself. A waveguide is shown with and without the heater layer in Fig. 9. A completed third order filter is shown in Fig. 10 (a) and a wafer diced into 3.5 mm columns, each holding 5 third-order filters, is shown in Fig. 10 (b).

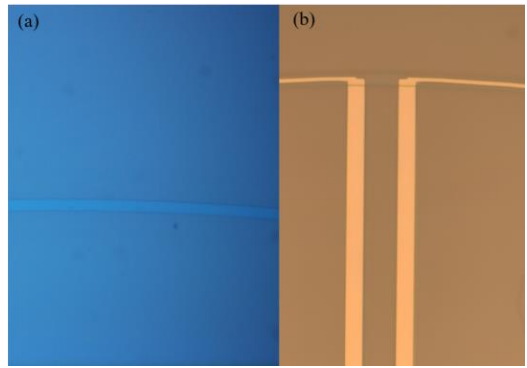

Fig. 9.    (a) $Si_3N_4$ waveguide with upper cladding deposited. (b) The metal layer over the waveguide layer is 5µm wide for the interconnect and 1.2µm wide over the waveguide.

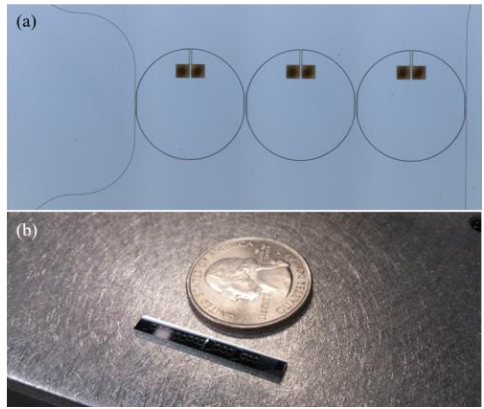

Fig. 10.   (a) Photo micrograph of the third-order filter.  (b) Image of a 3.5mm wide bar of 5 third-order filters relative to a quarter.

## 6   Characterization

A wavelength swept laser source was used to measure the passband and align and tune the rings. However, this approach limits the measurement of the filter stop band to the ER of the laser being used to test. In order to measure a stopband ER greater than 70 dB, an Agilent 86140B optical spectrum analyzer with sensitivity of -90 dBm used in combination with the tunable laser and EDFA as shown schematically in Fig. 11. Filter tuning and fiber coupling were achieved using a probe setup and precision fiber aligners as shown in Fig. 12 with the filter maintained at 20 degrees C using a TEC controlled stage.

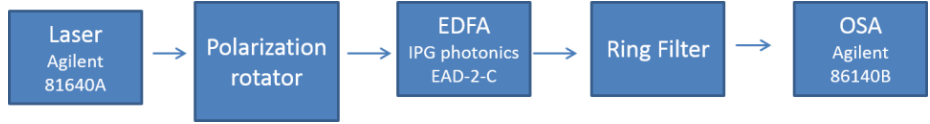

Fig. 11.   Schematic representation of measurement setup.

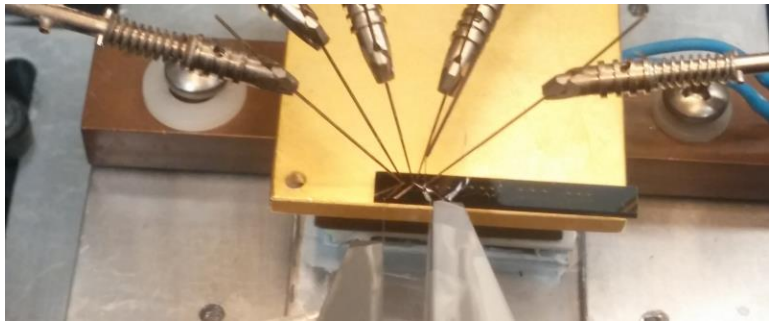

Fig. 12.   Photograph of third-order filter being measured.

### 6.1   *Waveguide Characterization*

A spiral structure, 0.5m in length, was fabricated to measure the propagation losses. Propagation losses were measured using an optical backscatter reflectometer (OBR) as described in [12].  Fig. 13 (a) is a TM polarized OBR trace showing reflected power as a function of propagation distance.  Fitting a slope to the trace in Fig. 14 (a) yields the waveguide loss, shown in Fig. 13 (b) for both TE and TM polarizations.  The two different modes have nearly the same propagation loss, with a minimum loss of 9.2dB/m and 10.5dB/m at 1590, and a loss of 15.1dB/m and 17.0dB/m at 1550.

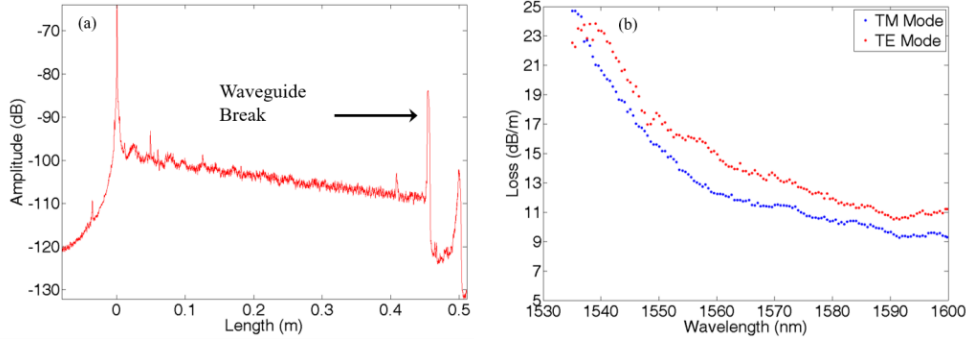

Fig. 13.  (a) OBR trace and (b) its resulting loss fit.  The trace is for the TM mode.

*6.2    Third-Order Ring Filter Performance*

Each ring within the filter is fabricated with an independently controllable platinum heater.  Due to small variations in individual rings within the filter, tuning is required to properly align the resonances as shown in Fig. 14 (a) and enables optimization of both the stopband and the passband as shown in Fig. 14 (b).  Filter tuning is achieved through small heater changes as the filter transmission is measured, a technique that has been automated for up to fifth-order filters as reported in [22].  The shape factor and ripple are set by the coupling ratio, as given in Eq. (1). Using this aligning technique, we found device yields greater than 90% across a single wafer.  Tuning the filter over its FSR is described in section 7.5.

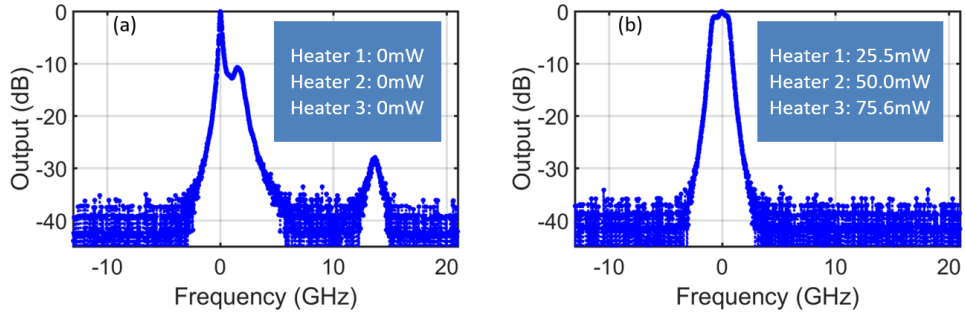

Fig. 14.    Wavelength sweeps of third-order filters.  The measurement is limited by photodetector dynamic range.  (a) shows a filter initially out of resonance, (b) shows the same filter tuned to resonance.

Relative optical power transmission of the third-order ring filter is shown in Fig. 15 by plotting the ratio of the input to output filter power, shown in Fig. 11.  The filter extinction ratio is measured to be 80dB as shown in Fig. 15 (a).  Fitting these values to equation (1) gives coupling values of $\kappa_1$=0.125 and $\kappa_2$=0.005, very close to the targeted values of 0.13 and 0.006 respectively.  The filter 3dB bandwidth and 20dB bandwidth were measured to be 1.60GHz and 3.12GHz respectively.  The filter input loss was measured using a laser set to the passband of the filter and received by a photodetector.  The power measured at the facet was 5.6dBm, and the power at the detector was 11.7dBm.  The average coupling loss measured on straight waveguide test structures was 2.4dB.  Removing the coupling loss from the power loss in the filter gives an insertion loss of 1.3dB.

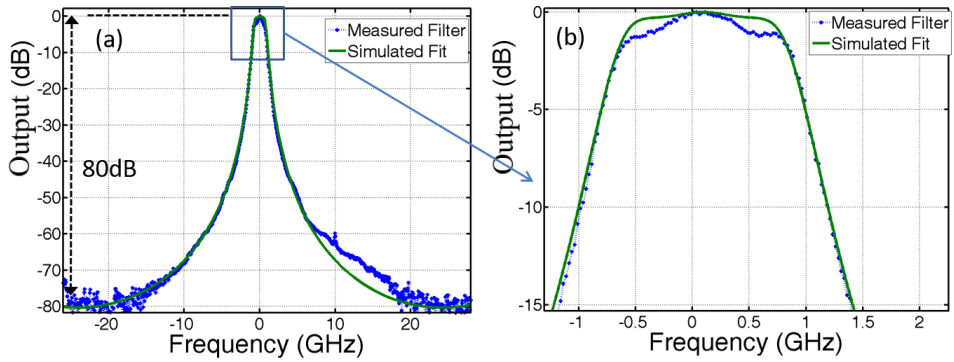

Fig. 15. (a) Third-order filter function, with an extinction ratio of 80dB and FSR 48.2GHz. The analytical fit yields $\kappa_1$ fit=0.125, $\kappa_2$ fit=0.005. (b) Third-order filter passband with a shape factor of .437 and no ripple.

### 6.3 Filter Tuning and Metal Absorption

In section 7.2 the heaters were used to independently align the rings to realize a third-order filter. If the power dissipated in the heaters is increased uniformly, such that the differences in power between each heater from the alignment are maintained, the filter can be tuned over its full FSR while maintain the filter shape. Tuning the rings in this manner results in an efficiency of 0.461GHz per mW of power per ring, equivalent to 0.105 W/FSR, shown in Fig. 16.

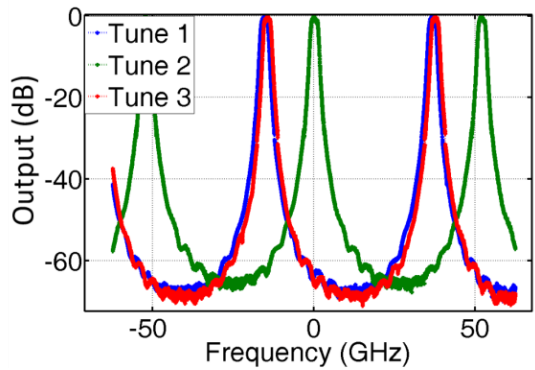

Fig. 16. A third-order ring filter is tuned over its full FSR. Tune 1 represents no thermal tuning, tune 2 represents 50mW of thermal tuning, and tune 3 represents 110mW of tuning.

The spiral test structure does not include a metal layer, and therefore the measurement does not include any loss incurred by the metal layer. To evaluate metal layer induced losses, we compare the losses of two identical first-order rings, one with a metal tuning layer deposited and the other without a metal tuning layer, shown in Fig. 17. Fitting the two filter functions to the theoretical model, we find the additional loss of the metal layer to be 1.7dB/m at 1550nm.

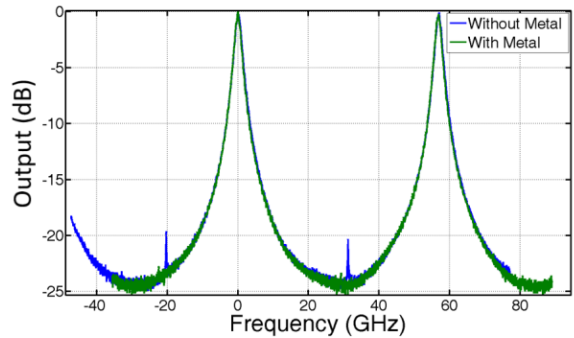

Fig. 17. First order rings with and without a metal layer are compared. Fitting the two curves to the theoretical model yields and additional loss, due to the metal layer of, 1.7dB/m.

## 7 Summary and Conclusion

We have presented the design, fabrication, and demonstration of a record high extinction ratio continuously tunable third-order ring filter. The measured 80dB ER, is a 100x improvement over the ER of previously reported $3^{rd}$ and $6^{th}$ filters, with a shape factor of 0.44. The individual third-order filters have FSRs and tuning ranges of 48.2GHz over 100% of the FSR with the ER, loss and shape factor maintained. The demonstrated performance for $3^{rd}$ order filters improves yield by requiring coupling of three rings as compared to 11-ring coupling.

Implementation in the ultra-low loss $Si_3N_4$ waveguide platform lends to monolithic integration with a wide range of both active and passive components that have been demonstrated including $Si_3N_4$-core with co-doped $AL_2O_3$:$Er^{3+}$ doped narrow linewidth WDM sources and thermal switches and delay lines. The demonstrated 80dB ER is critical for new applications that involve separation of closely spaced signals, for example the 50dB ER at 12GHz is ideal for separating pump and probe for Brillouin scattering. The filter characteristics are desirable for filtering of idler signals in FWM in microresonators for non-magnetic optical isolation and quantum communications and computing that employ frequency conversion and require separation of quantum entangled states for transmission.

These filters utilized the TM mode of high aspect ratio $Si_3N_4$ waveguides, due to initially high TE mode loss. Improved fabrication techniques have reduced the TE loss to that of TM loss. As future work, these filters could utilize the TE mode to increase the FSR by nearly a factor of two, while reducing the round-trip loss. The demonstrated filters utilize electrical heaters as a tuning source. Future work will involve alternative tuning techniques to improve the tuning speed and power dissipation.

## 8 Funding
Keysight Technologies

## 9 Acknowledgements
We thank Michael Davenport and Michael Belt for assistance with clean room fabrication. We also thank Daryl Spencer for discussions on ring resonators.